\documentclass[iop]{emulateapj}
\usepackage{epsfig}
\def\apj{ ApJ}
\def\aap{ A\&A}
\def\mnras{MNRAS}

\def\apjl{ ApJL}

\shorttitle{Mass and fate of the post-CE remnant}
\shortauthors{Ivanova }

\begin{document}

\title{Common envelope: on the mass and the fate of the remnant.}
\author{N.\ Ivanova$^1$
}
\altaffiltext{1}{University of Alberta, Dept. of Physics, 11322-89 Ave, Edmonton, AB, T6G 2E7, Canada}

\begin{abstract}{
One of the most important and uncertain stages in the binary evolution is the common envelope (CE) event.
Significant attention has been devoted in the literature so far to the energy balance 
during the CE event, expected to determine the outcome.
However this question is intrinsically coupled with the problem of what is left from the donor star
after the CE and its immediate evolution.
In this paper we argue that an important stage has been overlooked: post-CE remnant 
thermal readjustment phase.
We propose a methodology for unambiguously defining the post-CE remnant mass
after it has been thermally readjusted,
namely by calling the core boundary the radius in the hydrogen shell corresponding to the local 
maximum of the sonic velocity. 
We argue that the important consequences of the thermal readjustment phase are: (i) a change 
in the energy budget requirement for the CE binaries and 
(ii) a companion spin-up and chemical enrichment, as a result of the mass transfer that occurs during the remnant thermal readjustment (TR).
More CE binaries are expected to merge. If the companion is a neutron star, it will be mildly recycled during
the TR phase. The mass transfer during the TR phase is much stronger than the accretion rate during the common envelope,
and therefore satisfies the condition for a hypercritical accretion better. 
We also argue that the TR phase is responsible for a production of mildly recycled pulsars in double neutron stars.
}
\end{abstract}


\section{Uncertainty in the common envelope theory} 

In the standard treatment of common envelope (CE) outcomes via the ``energy formalism'' \citep{Webbink84},
the final separation of the binary is determined by equating  the binding energy of 
the (shunned) envelope $E_{\rm bind}$ to the decrease in the orbital energy $E_{\rm orb}$:

\begin{equation}
E_{\rm bind} = E_{\rm orb,i} - E_{\rm orb,f} = -\frac{ G m_1 m_2} {2 a_{\rm i}} + \frac{ G m_{\rm c} m_2} {2 a_{\rm f}} 
\end{equation}
Here  $a_{\rm i}$ and $ a_{\rm f}$ are the initial and final binary separations, $m_1$ and $m_2$ are the initial star masses
and $ m_{\rm c}$ is the final mass of the star that lost its envelope.

$E_{\rm bind}$ is considered to be the sum of the potential energy of the envelope and its internal energy,
and can be found directly from stellar structure for any accepted core mass
\cite[there are also modifications for $E_{\rm bind}$, where ionization energy or enhanced winds 
are taken into account, e.g.][]{Han95, Han02, Soker04}:

\begin{equation}
E_{\rm bind} = \int_{\rm core}^{\rm surface} \epsilon (m) dm  = \frac {G m_1 m_{\rm e}} {\lambda R_1} 
\end{equation}
Here  $\lambda$ is a parameter introduced to fit $E_{\rm bind}$; it characterises the donor envelope central concentration.
$m_{\rm e}$ is the mass of the removed giant envelope and is commonly assumed to be $m_{\rm e}=m_1 - m_{\rm c}$,
 $R_1$ is the radius of the giant star at the onset of CE, 
and  $\epsilon$ is the sum of the specific internal and potential energies. 

For the final balance  of energy, one more parameter is introduced, $\alpha_{\rm CE}$, to  
measure the energy transfer efficiency from the orbital energy 
into envelope expansion:

\begin{equation}
\alpha_{\rm CE} {\lambda} \left ( \frac{ G m_{\rm c} m_2} {2 a_{\rm f}} -\frac{ G m_1 m_2} {2 a_{\rm i}} \right ) =
\frac {G m_1 m_{\rm e}} {R_1}
\label{allam}
\end{equation}

We anticipate that introduction of the two parameters introduced accordingly two uncertainties.
It is common to remove these uncertainties {\it at the same time}, considering the product of $\alpha_{\rm CE}$ and $\lambda$, by means
of comparison of observations with the binary population synthesis calculations,
where the product of the two parameters is varied to match the observations.
However, this approach has shown inconsistencies with the observations, especially large for the formation rates of 
black hole LMXBs \citep{podsi03,Justham06}. In particular, for LMXBs this required $\alpha_{\rm CE}\lambda \ga 2$ \citep{Yun06}, although in massive giants
$\lambda \ll 0.1$  \citep{podsi03}, and $\alpha_{\rm CE}$ is bound to be $\le 1$.

The other way to reduce uncertainties is to consider them separately, 
e.g.\ one can try to determine an `accurate' value of $\lambda$ from stellar structure calculations.
It is then crucial to be precise about the definition of the core -- should only the hydrogen envelope be removed, 
or together with the H-burning shell, and so on \citep{Dewi01}. 
Without knowing what exactly counts as the core and which material ought to be ejected, the inferred $\lambda$ 
can vary by a factor of several {\it from this uncertainty alone}.

The physical reason for this variation is that in giants, within the hydrogen shell, the potential is strongly 
increasing towards the core.
The uncertainty increases as the mass of the donor increases, 
and changes from a about  a factor of 2 in intermediate mass stars at early giant stage 
to a factor of 20 and more for well-evolved massive stars \citep{Dewi01}.

We stress that neither observations nor theory provide now
a strong constraint on what post-CE remnant mass should be  
at the moment when the {\it dynamical} phase of the ejection ends.
It does not have to be the same as the mass of the remnant that we observe now, e.g., in double white-dwarf (WD) systems
or in sub-dwarf B stars: some remaining post-CE hydrogen-rich material can be easily removed through
strong winds similar to those on horizontal branch, or asymptotic giant branch, or in Wolf-Rayet stars etc. 
Between the dynamical phase and long-term evolution, the core will 
readjust itself on a thermal time-scale, and this has not been addressed.
Here, we address the problem of what the post-ejection mass could be,  
different regimes in which a post-CE remnant can shed its remaining hydrogen-rich mass and the consequences for a companion
due to post-CE mass transfer.

\section{The post-ejection remnant}

\subsection{The divergence point}

In SPH simulations of physical {\it collisions} between a RG and a neutron star (NS) it has been found
that not all hydrogen material is ejected along with the envelope -- a tiny layer 
of hydrogen, from the H-burning shell, remains \citep{Lombardi06}. This event 
is not directly comparable to a typical CE event in a binary as   
at the time of initial approach, at periastron, the H-burning shell of the donor could have been 
in the immediate Roche lobe of the intruder. This magnifies the mass-loss from the H-burning shell
and as such can decrease the mass of the post-CE remnant 
compared to a typical CE, where this shell might never be in the Roche lobe of the spiraling-in
companion. This example makes clear that even in a 
dynamical CE some hydrogen-rich material always remains.

On the other hand, studies of the evolution of stripped cores of low-mass RGs, have shown that
there is a minimum ``envelope'' mass $\delta m_{\rm e, min}$ that has to be left on the core in order
for the star to reexpand; if less mass is left on the core the star will contract and become a WD 
\citep{DS70}. This expansion or contraction of the remaining shell occurs on the thermal timescale
of remaining layer, $\tau_{\rm th}$.

It is plausible therefore to suppose that there is a unique ``divergence'' point $m_{\rm d}$ inside the hydrogen burning shell,
such that if a post-CE star has any mass above this point,
the star will continue to expand on $\tau_{\rm th}$. If its final mass is less than $m_{\rm d}$,
the star will shrink, also on its $\tau_{\rm th}$.
We recognise that  $\tau_{\rm th}$ might mean different values in the case of degenerate core (applicable
only to the remaining shell) or non-degenerate core (where it likely to depend on the core conditions).
We expect that the material above the divergence point, if left, 
will expand, in order to obtain thermal equilibrium, but is not required to escape to infinity without an additional 
energy source (such as the orbital energy). During this thermal readjustment (TR), it may also fill its Roche lobe.

\subsection{Calculations}

We tested this idea of ``divergence'' point on giants of several initial masses ($1,2,10,20,30\ M_\odot$).
The stars were evolved using the stellar code and input physics described in \cite{IT04}.
This code is capable of performing both hydrostatic and hydrodynamic stellar evolution calculations.
For a Roche lobe overflow evolution in binaries, it finds mass loss rates implicitly. 
Massive stars, where wind loss are important,  were evolved
with wind loss rates according to \citet{Vink01}, or,  where Vink rates are not applicable, 
according to \cite{Reimers78}.

For each initial mass, we chose 2-4 evolutionary states within the giant stage with different hydrogen-exhausted core masses $m_{\rm X}$.
As during the advanced evolution stages stars can shrink, we ensured 
that the chosen giants had expanded to their current radius for the first time.
On these giants, we imposed very fast (``adiabatic'') mass loss, $1 {M_\odot}$/year.
Such timescale for the mass-loss  $\tau_{\rm ML}$ -- about several initial binary orbits -- is comparable 
with a fast CE event. The lower bound on $\tau_{\rm ML}\ga1{\rm yr}$ should be clear as a CE event has to happen over at least one binary 
period at the initial Roche lobe overflow.

We do not imply that a CE ejection features a constant fast mass loss, and also do not study
the reaction of the outer (convective) envelope. We are interested in the reaction of the inner layers, 
which are most likely to remain after the envelope ejection has occurred.

For each mass coordinate, let us compare $\tau_{\rm ML}$ with the local 
thermal timescale $\tau_{\rm TH} (m) = E_{\rm bind}(m)/L(m)$  and
the local dynamical timescale $\tau_{\rm dyn}(m)$:
The mass-loss and the star evolution will be adiabatic if  $\tau_{\rm ML}\ll\tau_{\rm TH} (m)$.
The evolution can be described by hydrostatic approximation if $\tau_{\rm ML}\gg\tau_{\rm dyn}(m)$,
as a star will always acquire its hydrostatic equilibrium within a dynamical time, 
and its state at the hydrostatic equilibrium is defined by its thermal structure.

\begin{figure}[t]
  \includegraphics[height=.36\textheight]{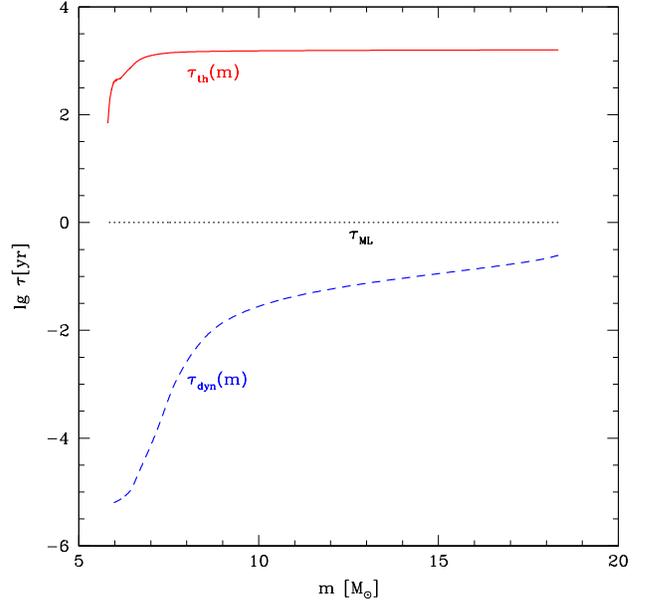}
  \caption{Comparison of the local thermal time-scale $\tau_{\rm th}(m)$ and the local dynamical time-scale $\tau_{\rm dyn}(m)$ with the mass-loss time-scale
$\tau_{\rm ML}$. Shown in the case of $18.5 M_\odot$ (ZAMS mass $20 M_\odot$, considered when  $R=750\ R_\odot$). 
\label{timescale}
}
\end{figure}

For most stars, $\tau_{\rm ML}$
is much shorter than any local thermal timescale (see Fig.~\ref{timescale}).  We note however that in the 
inner layers that are close to the cores of our most massive stars (20 and 30 $M_\odot$), 
the complete mass-loss sequence can take up to 10\% of the local thermal timescale of a few hundred years.
Thus, even such a fast mass-loss produces only approximately adiabatic evolution: 
some thermal evolution proceeds and is expected to be responsible, in particular, for some expansion  
of inner non-degenerate layers during the CE phase.
As a sanity check, we calculated additional mass-loss sequences for massive stars, with faster and slower mass-loss rates, 
and found only minor differences in the region of interest between the runs with $0.1, 1$ and $10~M_\odot {\rm yr}^{-1}$. 

On the other hand, local dynamical time-scales are longest at the surface and significantly shorter for innermost layers (Fig.~\ref{timescale}).
$\tau_{\rm dyn}(m)$ is comparable by the order of magnitude to  $\tau_{\rm ML}$ in the outer layer of the massive giants.
$\tau_{\rm dyn}(m)$ is however by 3 or more orders of magnitude smaller than  $\tau_{\rm ML}$ 
in the Helium rich layers, and closer to Hydrogen exhausted core, it is $\sim 10^{-5}\tau_{\rm ML}$, even in our most massive considered stars.
Therefore, although the evolution of the outer layers is indeed dependent on the inclusion of hydrodynamical terms,
the inner layers always have enough time to regain hydrostatic equilibrium, and are therefore insensitive under the adopted mass-loss rate. 
In summary, we find that for studies of the thermal reaction of inner layers that will form the remnant after the fast envelope ejection, 
the hydrostatic version of the code is sufficient.

As a result of mass loss evolution, we obtained sequences of (post-CE) remnants with different final (post-CE) masses, 
each of which then was evolved  for several $\tau_{\rm th}$, to check if this post-CE star is expanding or contracting.
We note that in our code the value of post-CE  mass could be resolved 
no better than pre-CE resolution in hydrogen shell, this is specifically important for low-mass giants 
(e.g., we have about 50 mesh points per  $0.02 M_\odot$ H-shell 
in 2 $M_\odot$ RG with a $m_{X}=0.52 M_\odot$).

To summarize, we separate the CE event into two stages: one resulting in the envelope ejection, 
and the subsequent thermal readjustment of the remnant. The latter part is a distinct phase 
unless the spiral-in (including the ejection of the envelope) takes place 
on a time-scale comparable to the shortest thermal time scale, several hundred years; and has not been heretofore treated in the literature.

Finally, an admonishment is in order: several estimates exist for the CE duration, neither one of them can boast 
conclusive observational evidence or indeed self-consistency. E.g., a `slow' CE could last for 100 years and longer \citep{podsi01}.
We can not justify which CE evolution timescale is more appropriate, and this is not the purpose of this paper. 
We concentrate on the `fast' event, however, we see no reason why our results should not be applicable in the `slow' case:
the core reaction will be similar, albeit lagging by the time the ejection takes. We also note
that a 10-times slower loss rate did not produce a significant difference in our calculations.

\subsection{Degenerate cores}

Indeed,  as in previous studies, we found that every low-mass giant with 
a degenerate core has a unique divergence point $m_{\rm div}$ such that
if post-CE mass is less than  $m_{\rm div}$, it contracts on $\tau_{\rm th}$.
All post-CE remnant with masses above $m_{\rm div}$ expand, 
create new outer convective zone and keep expanding even after $\tau_{\rm th}$.

After locating $m_{\rm div}$, we analyzed pre-CE giants structure 
to find what characteristics these points had in initial giants, before the stripping began.
For this, all giants, including massive, were used.
We noticed that among all the giants, $m_{\rm div}$ could have initially a wide range of hydrogen content, $X=0.08-0.58$,
and so a criterion involving specific constant value of hydrogen abundance could not be satisfactory.
Similarly,  another criterion discussed in the literature -- the location where the energy generation rate is maximum \citep{Dewi01} ---
does not coincide with the divergence point.
We found that in the considered models $m_{\rm div}$ is close to the ``maximal compression point''
$m_{\rm cp}$, which is the mass zone with the maximum value of `compression' $P/\rho$ 
in the hydrogen shell. 

\subsection{Non-degenerate cores}

For massive giants, as previously, a post-CE remnant also has divergence point that corresponds to the minimum post-CE expansion of the remnant,
although the overall response is different from the case of the giants with degenerate cores:

\begin{itemize}
\item For all possible remnant masses with $m\la m_{\rm cp}$, the core slightly adiabatically 
expands during the fast adiabatic mass loss.
Once we stop the mass loss, it can very slightly (a few per cent) expand and then shrink dramatically, becoming smaller than it was before the CE.
\item For larger remnant masses, as previously, the convective envelope is re-formed, and the star remains as an extended giant 
for a while. 
The envelope can become larger than the pre-CE giant\footnote{Here, we can not fully separate the post-CE TR 
expansion  from  a normal stellar evolution along a giant branch,
however we find that a post-CE star obtains after the CE a large radius {\it faster} than it would have otherwise.}. 
\item For the intermediate range of remnant masses, above the $m_{\rm cp}$, but below the boundary where the convective zone
redevelops, the core experiences a pulse on $\sim \tau_{\rm th}$, 
being able to expand by up to few hundred times more than this mass had as a radius coordinate before the CE. 
After the pulse, the post-CE star shrinks significantly, also becoming smaller than prior the CE.
\end{itemize}

\begin{figure}[t]
  \includegraphics[height=.36\textheight]{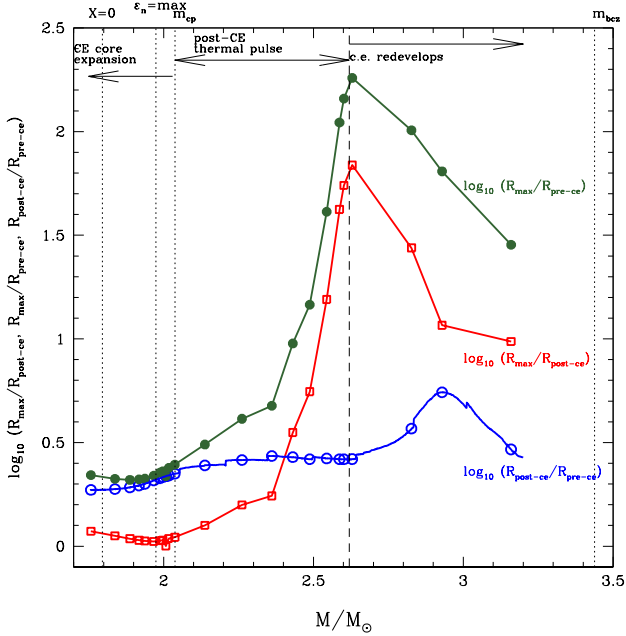}
  \caption{Size of a  post-CE remnant for a $9.75 M_\odot$ star (ZAMS mass $10 M_\odot$, considered when  $R=300\ R_\odot$). 
$R_{\rm pre-CE}$ is the radius coordinate of each considered remnant mass between the mass loss, $R_{\rm post-CE}$
is the radius of a post-CE star when the fast adiabatic mass loss stopped and $R_{\rm max}$ is the maximum
radius that this post-CE star had obtained within $\tau_{\rm th}$ (time to reach maximum ranged from 100 to 2,100 years).
Shown are the ratios of $R_{\rm post-CE}$ and $R_{\rm pre-CE}$ (blue line, open circles correspond to each calculated model),
$R_{\rm max}$ and $R_{\rm pre-CE}$ (green line, solid circles),
$R_{\rm max}$ and $R_{\rm post-CE}$ (red line, open squares).
$X=0$ is the mass of the hydrogen exhausted core, $\varepsilon_{\rm n}=max$ is where the nuclear burning has maximum energy generation rate,
$m_{\rm cp}$ is the ``compression point'' (where $P/\rho$ has the maximum value within the layer between $X=0$ and $m_{\rm bcz}$) and 
$m_{\rm bcz}$ is the mass coordinate of the bottom of the convective zone in  the pre-CE star.
\label{core_10_caseA}
}
\end{figure}

\begin{figure}[t]
  \includegraphics[height=.36\textheight]{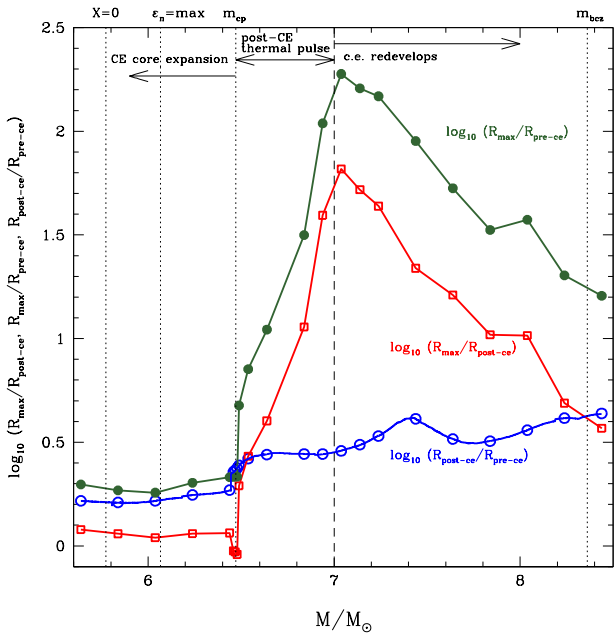}
  \caption{Size of a  post-CE remnant for a $18.5 M_\odot$ (ZAMS mass $20 M_\odot$, considered when  $R=750\ R_\odot$). 
Notations as in Figure \ref{core_10_caseA}.
\label{core_20_caseC}
}
\end{figure}

To illustrate these three types of the response, in 
Fig.~\ref{core_10_caseA} we show the typical case of $9.75 M_\odot$ (ZAMS mass $10 M_\odot$) star, taken when it had radius of $300\ R_\odot$.
For comparison, we show $18.5 M_\odot$ (ZAMS mass $20 M_\odot$) star with radius of $750\ R_\odot$
(see Fig.~\ref{core_20_caseC}). Even though this giant has profile of hydrogen  
qualitatively different from the considered above $10 M_\odot$ star, it shows similar behavior. 
The main difference with a $10\ M_\odot$ star is that there is a more contrasting response
between inside $m_{\rm cp}$ and outside it. 
In a $30 M_\odot$ giant this difference even stronger as
$m_{\rm cp}$ is located just below the bottom of hydrogen burning {\it convective} zone.

\subsection{The adiabatic response}

We recognize that the response we discuss above is non-adiabatic, but is the equilibrium response
(the one that a star experiences in order to obtain its thermal equilibrium).
Another important response to consider is the reaction of a star when the dynamical event ends, 
{\it true adiabatic} response. It is established that adiabatic response of the surface layers
depend on whether they are radiative or convective \citep{HW87,Soberman87}. In particular, radiative layers
on dynamical timescale tend to shrink, and convective layer remain the same or expand.
We checked for all considered model the location of the divergence point
and found that all of them are located within initially (pre-CE) radiative layers.
We conclude that immediate response for mass removal to the divergence point
is always a shrinkage and therefore does not affect our conclusions based on the thermal response.

\section{Consequences for the energy budget}

\begin{figure}[t]
  \includegraphics[height=.36\textheight]{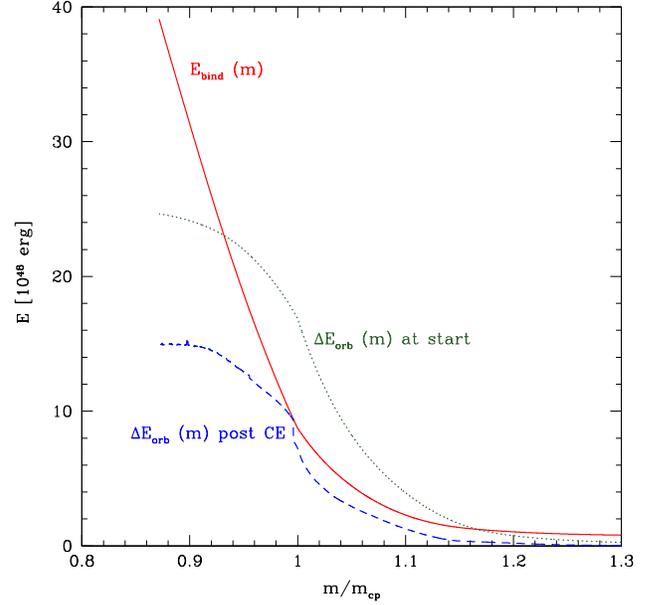}
  \caption{The initial binding energy (solid red line) and available orbital 
energies in the initial and final configurations (dotted green and 
dashed blue lines) for an $18.5~M_\odot$ star (ZAMS mass $20~M_\odot$) with $R=750 R_\odot$.  
When calculating the post-CE $\Delta E_{\rm orb}$, the sizes of the post-CE remnants 
for every given mass were used to define $a_{\rm f}$. 
The mass coordinate is normalized to $m_{\rm cp}$ (in this star $m_{\rm cp} = 6.47 M_\odot$).  
The $\Delta E_{\rm orb}$ are calculated assuming that the companion mass is $1.82~ M_{\odot}$.}
\label{fig_en_delta}
\end{figure}

As the core expands during semi-adiabatic mass-loss, a surviving binary must be {\it wider} 
when core expansion is taken into account.
Thus more binaries will merge \citep[this result is {\it opposite} to the claim
in][]{deloye10}. As an example, the giant in Fig.~\ref{core_20_caseC}  will easily survive a CE 
event with a NS of $1.4 M_\odot$ (with $\alpha_{\rm CE}=1$) if its core did not expand.
Setting core mass $\la 7.44 M_\odot$ will satisfy the energy budget to create a compact binary. 
However, if one takes into account the core expansion, the binary will merge:
for all core masses above $m_{\rm cp}$, the remnant will expand significantly and overfill its Roche lobe.
A minimum companion mass, for which both energy budget and post-CE core size are taken into account,
will be $\sim 1.82 M_\odot$ and the giant core in this case should have been removed to at 
least $m_{\rm cp}$ (see Fig.~\ref{fig_en_delta}).
It can be seen from this Figure that for a fixed companion mass, there is a unique solution where 
available orbital energy can exceed binding energy if the remnant 
expanded after mass loss, whereas a non-expanded remnant gives a wide range of the possible core masses, 
from 5.95 to 7.5 $M_{\odot}$.

\begin{figure}[t]
  \includegraphics[height=.36\textheight]{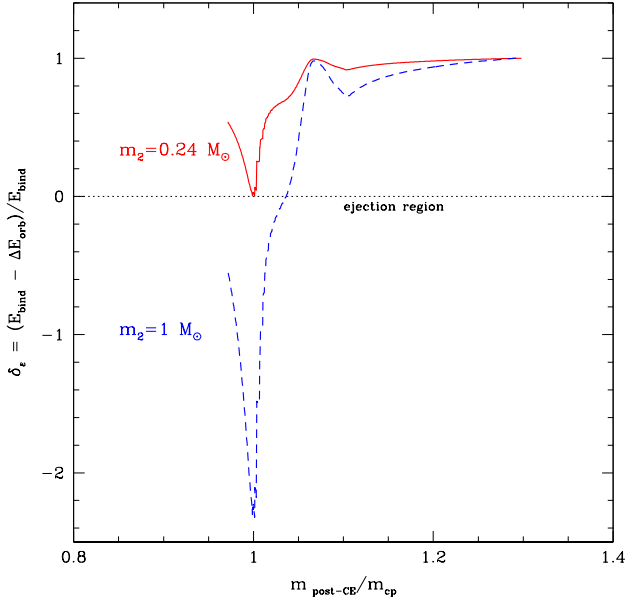}
  \caption{The energy expense in the the layers that can become a post-CE remnant for a $4.75~M_\odot$ star 
(ZAMS mass $10~ M_\odot$)  with $R=400 R_\odot$.
The two curves are for different companion masses (as marked). 
The mass coordinate is normalized to $m_{\rm cp}$.}
\label{fig_energy1}
\end{figure}

Let us introduce ``the energy expense'',  the difference between the 
required energy $E_{\rm bind}(m)$ and the available orbital energy $\delta E_{\rm orb}$, 
normalized per $E_{\rm bind}(m)$.

\begin{equation}
\delta_{\varepsilon} = \frac{ E_{\rm bind}(m) - (E_{\rm orb,i} - E_{\rm orb, f})}{E_{\rm bind}(m)}
\end{equation}
It is the (normalized) excess energy available to the envelope after all the matter above the given 
mass coordinate has been removed. Of course, a positive $\delta_{\varepsilon}$ signals that the removal process is not possible
from the energy considerations.
Fig.~\ref{fig_energy1} shows distribution of $\delta_{\varepsilon}$ as a function of the mass coordinate.
For these calculations, $E_{\rm orb, f}$ was assumed to be at the Roche lobe limited orbit for the post-CE remnant (semi-adiabatic
expansion is taken into account), so that available orbital energy is at its maximum.

Note that $m_{\rm cp}$ the energetically optimal position to remove the envelope to: 
it is the equilibrium point of the generalized force 
$\partial_m (E_{\rm bind}+E_{\rm orb})$. 
The shape and the location of the minimum of $\delta_{\varepsilon}$ only depends on the donor's 
energy profile up to the Roche radius, and does not depend on the companion mass. The latter only determines
the magnitude of the energy excess (see Fig.~\ref{fig_energy1}). 

Thus it is easy to determine the minimum mass of a companion which allows the survival of the binary:
as  it can be seen from the Fig.~\ref{fig_energy1}, it is such that will result in removal of mass to $m_{\rm cp}$, precisely. 

\begin{figure}[t]
  \includegraphics[height=.36\textheight]{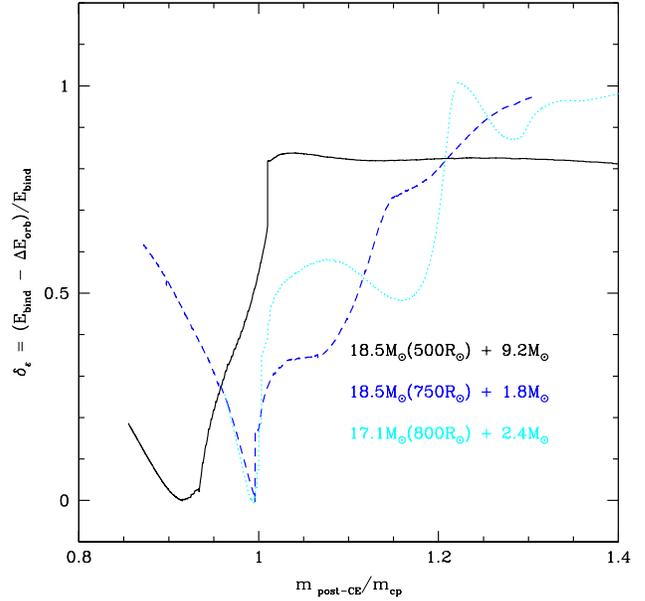}
  \caption{The energy expense in several giants with initial mass $20 M_\odot$. 
The mass coordinate is normalized to $m_{\rm cp}$.
Energy expenses are calculated using different assumed companion masses, to bring them to the same value for the energy minima.}
\label{fig_energy2}
\end{figure}

We verified that the coincidence of the minimum of the energy expense 
with shedding the envelope to about $m_{\rm cp}$ for minimum likely companion mass holds for many of the studied giants, 
though does not hold for giants that are early on the giant branch which only recently develop convective envelopes. 
E.g. in an early $10\ M_\odot$ we observe one more energy minimum, at a higher core mass $\sim 2.45 M_\odot$; 
the local energy minimum at $m_{\rm cp}$ nonetheless holds. 
In more massive early giants, the energy expense minimum is between $m_{\rm cp}$ and $m_{X}$  (Fig.~\ref{fig_energy2}),
its location is closer to the location of $\varepsilon_{\rm n}={\rm max}$ than to $m_{\rm cp}$;
the star still has the same reaction on expansion or contraction with respect to $m_{\rm cp}$ as other stars.

\section{Post-CE mass transfer}

\subsection{Consequences for the final post-CE mass}

Let us consider what happens if the companion was massive enough so that during CE not all mass to $m_{\rm cp}$
had to be removed. In this case a CE would end with a binary separation such that  
Roche Lobe overflow for a post-CE remnant during its TR will follow.
To consider this, we took a post-CE model showed in Fig.~\ref{core_20_caseC}, 
considering its remnant with $m_{\rm post-CE}= 6.84 M_\odot$ (larger than its $ m_{\rm cp} = 6.47\ M_\odot$).  
This post-CE remnant is then placed in a contact binary with an arbitrary companion of $3\ M_\odot$:
this mass self-consistently satisfies the energy required to shed the mass above $6.84 M_\odot$. 
For the mass transfer (MT) we can choose a fully conservative or fully non-conservative mode \footnote{Partial 
conservation, limited to Eddington rates can be considered as well, but with the MT rates that 
we find and describe, this case does not differ much with the fully non-conservative case.}. 

As expected, the post-CE remnant rapidly expanded and started the MT;
initially at a very high rate ($\sim 1-5 \times 10^{-2} \dot M_\odot$/yr), in accordance to $\tau_{\rm th}$ 
of the remaining hydrogen rich layer. 
After removing most of the layer above $m_{\rm cp}$, it 
slowed down to $\tau_{\rm th}$ of the core ($\sim ~10^{-4} \dot M_\odot$/yr). 
The MT continued till MT rates become comparable to the TR time-scale so that 
the core can shrink faster than it expands due to mass loss. 
In this particular example the core reached almost exactly the divergence point, 
shedding   $\sim 0.4 M_\odot$ during the MT so that the final mass was $m_{\rm cp}$. 
We also performed a MT calculation in the fully non-conservative regime. 
In this case the final mass of the core after the TR phase is the same as in the conservative calculations.

We also considered the case when the companion is a NS. 
We note that with a $20 M_\odot$ donor then, even for $\alpha_{\rm CE}=1$, 
the energy requirements for envelope ejection would not be satisfied 
if the core expands as much as we find after our fast mass loss (i.e. the binary would merge). 
The final mass of the remnant of the massive giant is the same,  $m_{\rm cp}$, 
as for the $3~M_\odot$ companion. 
If the mass transfer is fully conservative, the NS is presumably spun-up, 
as it would accumulate $0.34 M_\odot$. 
Again, in the case of a fully non- conservative regime, 
we find that the final mass of the post-CE remnant is $m_{\rm cp}$.

Next we considered a system with a $10\ M_\odot$ giant (same as shown in Fig.~\ref{core_10_caseA}), 
considering as the post-CE core $2.54~M_\odot$ ($m_{\rm cp} = 2.04$). 
In our MT simulations with a NS companion, less material above $m_{\rm cp}$ has been transferred,
only $0.24~M_\odot$, the final remnant mass is $2.3 ~M_\odot$. It might be connected to the fact that 10 $M_\odot$
star does not have such a sharp profile as a $20 M_\odot$ in the post-CE thermal pulse zone, 
and its post-CE expansion is more flatter until about this mass (see  Fig.~\ref{core_10_caseA}).
With a smaller companion mass, more of the post-CE remnant mass is stripped off.

\subsection{Consequences for the companion}

The MT rates that we encounter in the post-CE TR phase are highly super-Eddington and we face the obvious question
whether the MT is approximately conservative or almost non-conservative, as this is crucial
for a companion.
The question what happens if the mass-accretion rate on a NS exceeds 
Eddington limit has been discussed extensively in the literature, in particular, the regime in 
which it exceeds $\dot M_{\rm Edd}$ {\it by many orders of magnitude}.

\cite{1979MNRAS.187..237B} showed that if the accretion rate is extremely high, few times $10^{-4}M_\odot~{\rm yr}^{-1}$, 
then within some volume \citep[the ``trapping radius'', e.g.][]{1999ApJ...519L.169K} around the star the diffusion 
of photons outward cannot overcome the advection of photons inward. 
While a black hole can swallow all the material in this case, 
if the accretor is a NS, radiation pressure near the NS's surface resists 
inflow in excess of the Eddington limit, likely leading to creation of a Thorne-Z\'ytkov object.
\cite{Blondin1986} has also found that when MT rates exceed the Eddington rate by $10^{3}\times L_{\rm Edd}/c^2$ or more, 
the accretion proceeds in a hypercritical regime. 

Hypercritical accretion was then argued to be responsible for such efficient material accumulation during a CE event,
that a NS is likely to convert to a black hole \citep{1989ApJ...346..847C}. 
\cite{Brown1996} used this argument to understand double NS formation.
He showed that indeed in a CE event the Bondi-Hoyle-Lyttleton accretion rate is 
about $10^4 \times \dot M_{\rm Edd}$ and a NS can accumulate up to $1~M_\odot$. He argued that in this case,
considering that a number of the discovered double NS have masses closer to the lowest possible NS mass limit, 
a double NS can be formed only from a binary with almost similar
initial masses, evolving then via double CE event, before either of the NSs was formed.

\cite{1991ApJ...376..234H} considered neutrino losses during accretion on a NS. 
They studied in detail the regimes of the mass accretion  $10^{-4} \la \dot M \la 10^4 \, M_\odot {\rm yr}^{-1}$,
and found that radiation diffusion becomes important  
when the accretion rate falls below $10^{-3} M_\odot {\rm yr}^{-1}$, for smaller rates the radiation pressure
can not support an envelope around NS surface and can not cease the infall of the material.
We note that the accretion rate that separate the hypercritical accretion with the accretion when the
radiation diffusion dominates in this case
($10^{-3} M_\odot {\rm yr}^{-1}$) is higher than the one found to work during a CE event 
($2\times 10^{-4}  \dot M_{\odot} {\rm yr}^{-1}$).

If the latter estimate is more proper than in the studies listed above, then it is possible that 
a CE hyper-accretion, as having too low mass accretion rate,
does not lead to a significant accumulation of the material.
Post-CE core expansion, however, in either case can lead to a hyper-accretion regime, as during this thermal pulse
it provides a much higher mass accretion rate. This can lead to a NS spun up.
In our calculations, MT rates exceeded $10^{-3} M_\odot {\rm yr}^{-1}$ long enough to accrete in hypercritical regime on a NS 
$0.29 M_\odot$ in the case of $20 M_\odot$ giant and $0.09 M_\odot$ in the case of a $10 M_\odot$.
 
We note that the observed double NSs are generally mildly recycled, 
having periods  $0.024-2.7$ seconds \citep{{2004Sci...304..547S}}.
The mass distribution of those with mass measurement errors $\la 0.02M_\odot$ 
are such that the difference between the masses of the NSs 
is $\la 0.1M_\odot$ \citep[e.g., see data in ][]{2004Sci...304..547S,2010arXiv1011.4291K}.
Their location on the $P - \dot P$ diagram for galactic field NSs is also intermediate between the millisecond pulsars and
non-recycled ones \citep{1999ApJ...520..696A}, and the post-accretion period 
is likely to be not millisecond \citep{2005ASPC..328..113L}. 
It could be a sign that the very rapid mass transfer followed the CE
and preceding the second NS formation is not capable to fully spin up and efficiently reduce the magnetic field on 
a NS that was formed first. 
Same mechanism can lead to a formation of mildly recycled binary pulsars with low-mass
white dwarf companions \citep{Li2002, Del08}. 

\section{Conclusions}

We analyzed a set of giant models with respect to their likely post-CE response. 
Although our set was not exhaustive, it did exhibit a clear trend that allowed us to conclude that:
i) every giant has a well-defined post-CE remnant after it has been thermally readjusted, 
most likely given by the divergence point (see discussion below);
ii) the divergence points, at the current resolution, are best approximated by the 
point in the hydrogen burning shell that had maximal compression (local sonic velocity) $m_{\rm cp}$ prior to CE.
This definition allows us to find quickly a post-CE core mass for any giant without performing mass loss calculations.

We remark that this divergence point does not necessarily mark the 
final mass of the remnant  (e.g., the stellar wind in He rich stars 
could quickly and effectively remove the remaining hydrogen-rich envelope) or immediate
post-ejection mass, however it marks the mass after the thermal core readjustment.

The post-TR core, defined by $m_{\rm cp}$, most likely coincides with the post-ejection mass
in low-mass giants, where reestablishing of a convective envelope
for masses above the divergence point happens on a few dynamical timescales,
leading to another (likely unstable) MT event.
As the $E_{\rm bind}$ of remaining shells during this period has not been changed much
compared to the pre-CE state, the whole sequence of events can be considered as one CE event
with a final core being $m_{\rm cp}$.

For giants with non-degenerate cores the situation is more complicated. 
It is not possible to claim that the divergence point defines the post-CE core 
(dynamical phase) uniquely from the energetic budget point of view. 
However, the divergence point does appear to define the remaining post-TR core, 
if the post-CE configuration allows Roche lobe overflow for the post-CE remnant during the core TR. 
During this period, (stable) MT can proceed, resulting in the enrichment 
of a companion star with material from the hydrogen-burning shell. 

In all cases, the post-CE remnant of a giant with a non-degenerate core has 
expanded by few times by the end of TR. 
Remnant masses greater than $m_{\rm cp}$ lead to greater expansion; 
for each particular giant star, the minimum possible size change 
is for a remnant mass $\sim m_{\rm cp}$.  
The expansion of the remnant means that less orbital energy is available 
to eject the envelope than if there was no expansion, 
since the surviving binary must be correspondingly wider.  
This leads to a reduction in the number of binaries which can survive CE.  
So fast CE allows more binaries to survive the end of the dynamical phases. 
Slower, self-regulating CE leads to more mergers.

We note that for both types of giants $m_{\rm cp}$ firmly represents 
only a maximum post-TR core mass, as we can not fully rule out that 
the dynamical phase will not already have removed mass below $m_{\rm cp}$.   
However, we argue that such extra mass loss is not likely to happen if, 
during the final stages of the spiral-in, the characteristic orbital 
evolution time is comparable to the core response time near $m_{\rm cp}$ point, 
which is as short as 10-100 years.

We also find that in most evolved giants,
the energy required to shed the envelope down to $m_{\rm cp}$ is the minimum energy expense:
per total $E_{\rm bind}$ unit, it requires more orbital energy to remove either less or more of the mass from the expanded core.
It is fully reasonable to remove less of the envelope (and have a bigger post-CE mass) once
$E_{\rm bind} < \alpha E_{\rm orb}$, the TR phase will then remove mass down to $\sim m_{\rm cp}$.
However, it is not plausible to remove the envelope deeper than to $\sim m_{\rm cp}$:
for any remnant mass less than $m_{\rm cp}$, the 
difference between $E_{\rm bind}$ and $\alpha E_{\rm orb}$ increases compared to their value at  $m_{\rm cp}$.

We suggest therefore that a divergence point uniquely defines the core
in the post-TR phase, and then a slow wind loss
phase follows the CE with almost no core evolution\footnote{We do not include here evolution on nuclear timescale 
which will follow as usual; e.g., a core of a small enough mass can again become a giant, as it is customary for low-mass He stars.}
A companion can be spun up during the MT effected by the remnant's TR.
Roche lobe overflow phase is short, however can lead to at least mild recycling.
If the companion is a NS, its recycling will depend on whether 
the conservative mass transfer (due to the hypercritical accretion) is possible or not.
We find that hypercritical accretion is more likely during this TR phase than
during a CE, as the mass accretion rates are significantly higher.
It also leads to a smaller mass accumulation than is found to occur during a CE, corroborating
the mass distributions of the observed double NSs.
We conclude that the post-CE TR phase can be responsible for a formation of mildly recycled pulsars 
in post-CE binaries and specifically in double NSs.


\section{Acknowledgment}

NI thanks S.~Justham  and C.~Heinke for constructive comments and 
acknowledges support from NSERC and Canada Research Chairs Program.

\end{document}